\documentclass{cup-hpl}

\usepackage{lipsum}
\usepackage{etoolbox}
\usepackage{textgreek} 
\newcommand{\um}{\textmu m} 

\newcommand{\topow}[2][0]{\ifnumcomp{0}{=}{#1}{}{$#1~\cdot$~}$10^{#2}$}

\newcommand{\Wcm}[2][0]{\topow[#1]{#2}~W/cm$^2$} 

\usepackage{graphicx, float}
\usepackage{amsmath}
\usepackage{xcolor}

\usepackage[normalem]{ulem}

\begin{document}
\newtheorem{theorem}{Theorem}
\shorttitle{Direct Electron Acceleration by Radially Polarized Ultra-Intense Laser Focus During Ionization of High Charge States of Neon}                                   
\shortauthor{Hissi, Ngirmang,Smith and Chowdhury}

\title{Forward and Backward Electron Acceleration by Radially Polarized Ultra-Intense Laser Focus Seeded By Field Ionization of High Charge States of Neon}
\author[1]{Nour El Houda Hissi\corresp{
                       \email{hissinourelhouda@gmail.com}}}
\author[2]  {Gregory K. Ngirmang}
\author[3] {Joseph Smith} 
\author[1]{Enam A. Chowdhury}

\address[1]{Department of Material Sciences and Engineering, The Ohio State University, Columbus, Ohio, 43210, USA}
\address[2]{Singapore University of Technology and Design}
\address[3]{Department of Mathematical and Computational Sciences, The College of Wooster, Wooster, OH, USA}

\begin{abstract}
Thanks to the fabrication of large aperture phase optics, ultra-intense relativistic laser plasma interaction (RLPI) experiments with complex polarization states are becoming feasible. In this work, we perform a computational investigation of direct acceleration of electrons produced during ionization of underdense neon gas using a tightly focused and radially polarized Petawatt-class short pulse lasers by numerically solving the relativistically invariant Lorentz equations, incorporating semi-classical tunneling ionization and Monte Carlo type sampling of the focal volume. The accelerated electrons energy gain increases at longer laser wavelengths and GeV energies are reached for electrons ionized from the neon inner shells, which are field ionized near the peak of the pulse. Backward acceleration of electrons is observed for a range of initial positions and phases of ionization of neon charge states. This apparent counterintuitive phenomenon is directly linked to the radial polarization state of the incident laser beam that results in a strong longitudinal electric field Ez when tightly focused, where electrons ionized near the focal center at the phase when Ez is pointed toward the forward propagation direction experiences an initial push in the backward direction. A parametric study of the phenomenon by varying laser parameters is presented, and a 3D particle in cell (PIC) simulation is considered to confirm the existence of this phenomenon.

\end{abstract}


\maketitle

\section{Introduction}
Since the first laser was invented in 1960, the maximum intensity of lasers increased rapidly, pushing the peak power to the petawatt range \cite{Danson2015} and investigators believe that even higher peak powers may be reached in near future \cite{Danson2019}. Thus, Pettawatt-class lasers have stimulated considerable progress in the field of strong light and matter \cite{Salamin2006} interaction including laser driven particle acceleration. Over the years, different particle acceleration schemes were theoretically and experimentally investigated \cite{Hafizi1994,Sprangle1996} and beams with highly charged particles ranging from few MeV (mega electron-volt) \cite{Geddes2004,Goers2015} to GeV (giga electron-volt) \cite{Hogan2005,Wang2013} energies were reached. One of the most widely known particle acceleration scheme so far has been the Laser or plasma wakefield acceleration (LWFA/PWFA) \cite{Esarey2009,Guenot2017} where the excited large-amplitude plasma wave accelerates electrons while using a pulse with a weak or almost non-existent axial component of the electric field. Direct laser acceleration of particles (DLA) \cite{Salamin2010,Wong2017} also opens new horizons and offers promising avenue when intense ultra-short and tightly focused laser pulse is used. \cite{Payeur2012,Varin2013} 

In recent years, different laser polarizations including linear, elliptical and circular- have been investigated to achieve energetic electron beam \cite{Sohbatzadeh2009,Nanni2015}, but radially polarized (RP) pulses gained tremendous interest among all polarizations thanks to their unique geometrical properties which make them ideal for accelerating electrons to relativistic energies\cite{Zaim2017}. Their radial electric field and azimuthal magnetic field components help trapping electrons and confining them to a region close to the propagation axis while their axial electric field component directly accelerate them \cite{Zaim2017,Salamin06-2}. A variety of methods has been developed to focus RP pulses to micron size focal spot in order to achieve a strong longitudinal electric field for effective electron acceleration\cite{Dorn2003,Kaltenecker2016}. The capability of RP laser beam to accelerate electrons to GeV energies in vacuum was proposed using the lowest order RP fields of a focused and highly intense petawatt power laser beam\cite{Salamin06-2}. Wiggler magnetic field can also be used to boost electron energy and support their motion to retain the energy gain\cite{Kant2018}. GeV-electron beams were also achieved using charged ions as a source of electrons\cite{Salamin06}. Electron generation during ionization of low-density gases occurs in the semi-classical regime where electrons tunnel through a quasi-static barrier formed by the combined laser field and the Coulomb potential, from the bound to a free state. This process can be accurately modeled using semi-classical ionization model \cite{Smirnov1966,Perelomov1966}, and has been shown to be accurate up to $10^{21}$ $W/cm^2$ \cite{Chowdhury2001}. Ionization of high-Z materials has also been used to inject electrons in the acceleration phase for efficient acceleration of electrons  \cite{Hu2006,Singh2011}.

The purpose of this paper is to investigate the influence of the radially polarized laser wavelength on the direct electron acceleration and energy gain during the ionization of Ne$^{7+}$ and shed light on a new unconventional behavior of electrons accelerated under radial polarization; an electron can be pushed in the opposite direction to propagation under precise initial conditions. Laser wavelengths of 0.8 and 1 $\mu$m were chosen for their availability in current ultraintense laser systems, whereas 2 $\mu$m was chosen for future high repetition rate ultraintense laser platforms based on thulium:YAG, which recently have shown significant progress \cite{Tamer2021}. For linear and circular laser polarizations, electrons are only accelerated towards the forward (laser propagation) direction, due to the Lorentz force. Here our simulations show that with strong radially polarized laser fields, direct electron acceleration in the backward direction is possible. A Matlab code was built based on a combination of the Monte-Carlo method to generate particles randomly within a specific volume and the 5th order Runga-Kutta method with a controlled time step (attosecond range) to solve the relativistic invariant differential equations describing the electron dynamics and obtain the electron trajectory and momentum. We additionally present full-3D particle-in-cell (PIC) simulations to confirm the existence of backwards propagating electrons in the presence of a radially polarized short pulse laser by comparison to electrons accelerated by a linearly polarized short pulse laser of the same intensity, wavelength and pulse duration.

\section{Direct Forward and Backward Propagation}
\subsection{ \textbf{Electron Dynamics}}
\subsubsection{\textbf{Ionization :} }\hspace*{\fill} \\

In these simulations, we generate electrons from the interaction of high intensity laser pulse with Neon gas ions through the ionization process. As the interaction starts, the electrons are firstly removed from the outer shells of the atoms and thereafter from their inner shells. Ionization of low-density gases occurs specifically in the quasiclassical regime where the strong laser field distorts the Coulomb binding potential leading to tunneling electrons from a bound to a free state. An accurate tunneling model in this regime is the semiclassical model \cite{Esarey1993}. The numerical simulations used in this paper are valid in this regime. The electrons from the outer shells require a lower energy to be expelled compared to the electrons of the inner shell, they are therefore generated close to rising edge where electric field is relatively low and do not gain enough energy, while the electrons from inner shells are generated close to the peak of the laser pulse where electric field is relatively high and gain considerable amount of energy that can range from MeV to GeV with low angle of emittance and small energy spread \cite{Nanni2015}.

Due to the large difference between ionization potentials of the 8th and 9th electrons of neon, using a low intensity laser prepulse is necessary to remove the outer electrons and then apply the high intensity main pulse to accelerate the inner shell electrons to ultra-high energies \cite{Singh2011}. However, targeting the inner shell of neon requires a laser intensity in the order of $10^{21}$ $W/cm^2$ and up which mandates a more sophisticated laser infrastructure. In this study, we will be considering the difference between ionization potentials of the first and the 8th electrons of neon, which allows us to consider laser intensities in the order of $10^{19}$ $W/cm^2$ \cite{Palaniyappan2005} and higher, making an experimental verification of the theoretical results presented in this paper more accessible for current facilities.  Thus, we will assume that high field ionization from a pre-pulse is used before the main pulse to remove the seven first electrons of the outer shell (L) of Neon, while the main laser pulse directly interacts with the Ne$^{7+}$, Ne$^{8+}$ and Ne$^{9+}$. In this paper, we will only present and discuss the acceleration of electrons resulting from the interaction between Ne$^{7+}$ and the main pulse. We also keep the gas density sufficiently underdense, so that plasma effects can be neglected.

\begin{figure}
\centering
\colorbox{black!10}{\minipage[t][4.6cm][t]{8.3cm}
\hfill
\includegraphics[width=8.3cm,height=4.6cm]{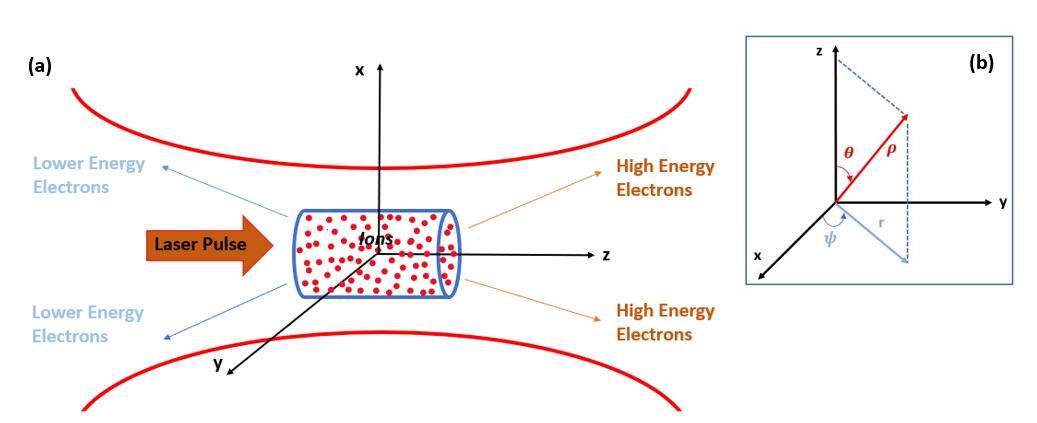}
\endminipage}
\caption{(a) Schematic of electron acceleration from ionization in a Gaussian laser pulse, (b) Representation of spherical polar coordinates used to localise an accelerated electron.}
\label{fig:Fig1}
\end{figure}

We used a Monte Carlo method to randomly generate $10^5$ Neon ions Ne$^{7+}$ considered initially at rest. The accelerated electrons have initially zero momentum and therefore zero velocity. We name $t_0$ the ionization time at which the electron is born. It strongly depends on laser intensity, the atomic number of the considered gas and its ionization potential. Fig.\ref{fig:Fig1} shows the schematic of electrons accelerating in both forward and backward directions during the ionization of Neon gas using a high intensity short pulse laser. The laser propagates along the z-axis and is focused at origin ($x= y=z=0$). Neon atoms were randomly distributed around the z-axis within a cylindrical volume with length equal to the Rayleigh range and radius equal to the considered wavelength. The particle density is kept uniform around the focal spot. The paraxial and non-paraxial electric and magnetic laser fields considered here are detailed in \cite{Salamin06,Varin2013}. Throughout the paper, the laser intensities mentioned  correspond to peak Gaussian laser focal intensities with corresponding fields, with the understanding that peak fields and corresponding intensities are lower for a radially polarized pulse in the same equivalent \textit{f}/\# system. Another important point to note is that the field and electron dynamical corrections due to non-paraxial fields for all cases presented here (\textit{f}/\# = 5) were calculated and compared to the results obtained using paraxial fields, and were found to be negligible. Therefore, the results presented here were chosen to be obtained by the paraxial field approximation. For shorter \textit{f}/\# systems, paraxial field approximations become more significant, and will be discussed in a later work.

\subsubsection{\textbf{Acceleration:} }\hspace*{\fill} \\

In the paraxial limit, the lowest order of an RP laser beam (TEM$_{01}$) is characterized by its central dark intensity region that turns into a bright spot of sub-wavelength diameter under tight focusing conditions \cite{Varin2013}. This particular behavior is due to the beam symmetry that favors a strong longitudinal electric ﬁeld component making it well-suited for the task of accelerating electrons to relativistic energies. For large laser focal spot radii, the incident laser pulse is assumed radially polarized and propagates along the \textit{z-}axis with electric field ($E$=r $E_r$+z$E_z$) where \cite{Varin2013}:
\begin{equation}
E_r = E_0\frac{r}{r_0f^2}cos(\phi)exp\left(\frac{-r^2}{r_0^2f^2}\right)exp\left(\frac{\left(t-\frac{z}{c}\right)^2}{\tau_0^2}\right)
\end{equation}
\begin{equation} \label{eq2}
\begin{split}
E_z& = \frac{2E_0}{k_0r_0f^2}\left(\left(1-\frac{r^2}{r_0^2f^2}\right)sin(\phi)-\frac{zr^2}{z_rr_0^2f^2}cos(\phi)\right)\\
&exp\left(\frac{-r^2}{r_0^2f^2}\right)exp\left(\frac{\left(t-\frac{z}{c}\right)^2}{\tau_0^2}\right)
\end{split}
\end{equation}

where \(\phi=k_0z-\omega_0t-2 \arctan{z/z_r}+zr^2/z_rr_0^2f^2+\phi_0\), \(f^2=1+\left(z/z_r\right)^2\), \(k_0=\omega_0/c\), \(z_r=k_0r_0^2/2\) is the Rayleigh length, $r_0$ is the minimum laser spot radius, $\omega_0$ is the laser frequency,$\tau_0$ is the laser pulse duration and c is the speed of light in vacuum. The magnetic field component $B_\theta$ related to the laser pulse can easily be deduced from Maxwell’s equations. The paraxial approximation leads to ever increasing inaccuracies in fields for the case of small f numbers 
\cite{Maltsev2003, Chowdhury2004}, a 5th order correction of electromagnetic fields of the laser is therefore implemented here to correctly describe electron dynamics and avoid erroneous results in electron trajectories and energies \cite{Salamin06}. Throughout this paper, time, length, velocity, momentum and laser electric and magnetic fields have been calculated in SI units. 

\begin{figure*}[hbt!]
\centering
\includegraphics[width=16cm,height=10cm]{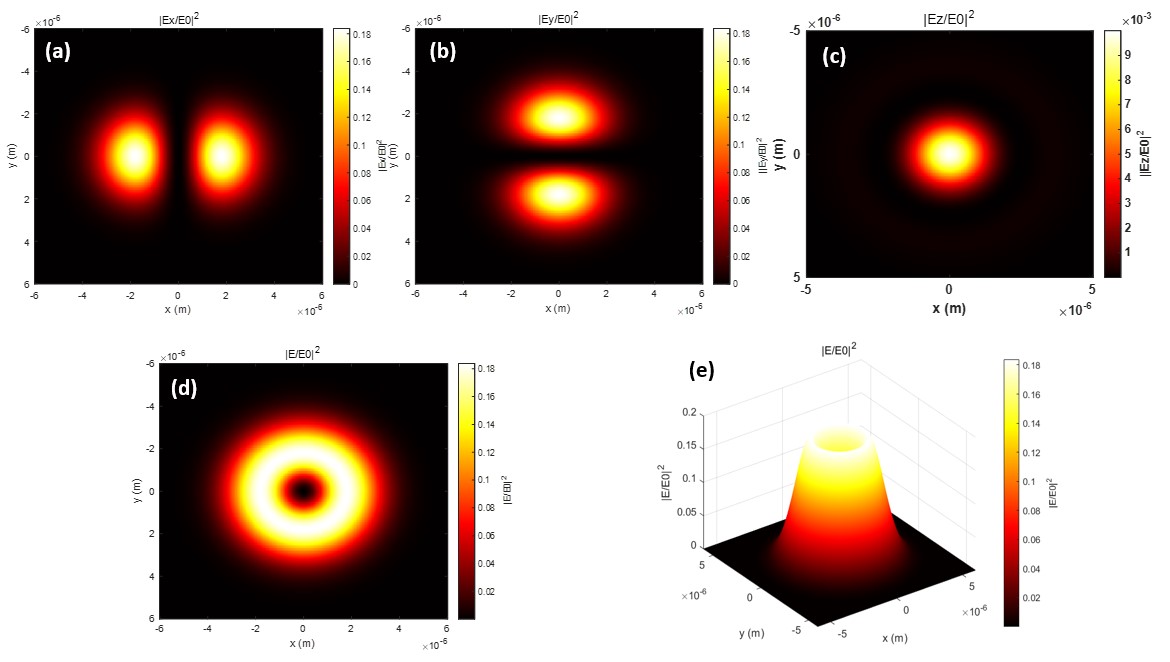}
\caption{\textsf{Distribution of normalized squared electric fields in x-y plane obtained for an \textit{f}/\# = 5 at an intensity $I$ = \Wcm[5]{19}. $E_0$=$1.94.10^{13}$ V/m is the incident electric field strength. (a) Normalized squared $\lvert E_x \rvert $, (b) Normalized squared $|E_y|$, (c) Normalized squared $|E_z|$, (d) Normalized overall distribution of the electric field  $E=\sqrt{|E_x |^2+|E_y |^2+|E_z |^2}$, (e) 3D plot of the normalized overall distribution of the electric field sliced with a plane through the center. Please note that in this case, radially polarized pulses can achieve only $\sim$18\% of the peak focal intensity of the Gaussian Laser focal intensity mentioned above.}}
\label{fig:fig2}
\end{figure*}

\begin{figure*}[hbt!]
\centering
\includegraphics[width=16cm,height=10cm]{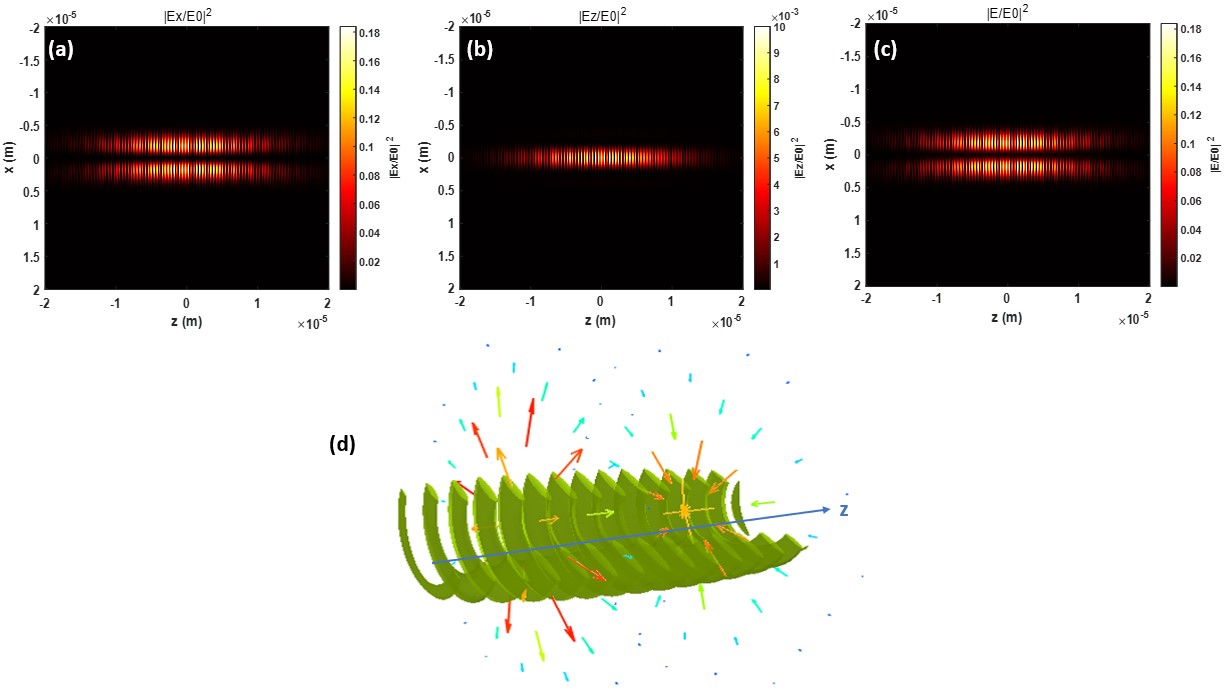}
\caption{\textsf{Distribution of normalized squared electric fields in x-z plane obtained for an \textit{f}/\# = 5 at an intensity $I$ = \Wcm[5]{19}. (a) Normalized squared $|E_x|$, (b) Normalized squared $|E_z|$, (c) Normalized overall distribution of the electric field  $E=\sqrt{|E_x |^2+|E_y |^2+|E_z |^2}$, (d) 3D plot of the normalized overall distribution of the electric field.}}
\label{fig:fig3}
\end{figure*}

Fig.\ref{fig:fig2} (d) shows the doughnut-shape profile of the focal spot. In this case, most of the electric field energy is concentrated in its transverse component $E_r$=$\sqrt{|E_x|^2+|E_y|^2}$, resulting in $|E^2|$$\approx$$ |E_r^2| $ as shown in Fig.\ref{fig:fig2} (b) and (c). $E_x$ and $E_y$ represent the projection of the radial component of the electric field $E_r$ in x and y directions.

Electron dynamics in a strongly relativistic electromagnetic field is significantly altered by the radiation reaction (RR) force also called radiation friction\cite{Zhang2018}, meaning that the electron, while being accelerated, emits electromagnetic radiation leading  to significant energy and momentum losses. To describe the interaction between the incident laser pulse and the considered under-dense neon gas, we use the Newton-Lorentz equation incorporating the Landau-Lifshitz form of the radiation reaction (RR) force \cite{Bulanov2011}:
\begin{equation}
\frac{d \textbf{p}}{d t}=- e(\textbf{E}+\textbf{v}\wedge \textbf{B})-\frac{2e^4 \gamma^2}{3m c^5} \textbf{v}((\textbf{E}+ (\textbf{v}\wedge\textbf{B}))^2-(\frac{1}{c}\textbf{v}.\textbf{E})^2)
\end{equation}

\textbf{E} and \textbf{B} are the laser's electric and magnetic fields, while \textbf{p},\textbf{v}, \textbf{m} and \textbf{-e} are respectively the electron's momentum, velocity, mass and charge.
The importance of quantum radiation reaction is characterized by the invariant quantum nonlinear parameter given as follows \cite{Blackburn2020,Di Piazza2010}:
\begin{equation}
\chi = \frac{e E_0\gamma_0(1+\beta_0)}{m^2}
\end{equation}
where $\gamma_0$ and $\beta_0$ are respectively the electron's initial velocity and initial lorentz factor. In our model, the particles are considered to be initially at rest, which means that their initial velocity and energy are both equal to zero, and thus equation (5) is simplified to :
\begin{equation}
\chi = \frac{e}{m^2}E_0
\end{equation}
where $m^2/e$ is the critical field of QED \cite{Blackburn2020} that marks the threshold for nonperturbative electron-positron pair production from vacuum and is equal to $E_c = 1.326 \times 10^{18}$ V/m. If we consider the incident laser intensity to be $I$ = \Wcm[5]{21}, the corresponding incident electric field strength is $E_0 = 1.94\cdot 10^{14}$ V/m and the quantum nonlinear parameter in this situation is equal to $1.463\cdot 10^{-4}$ meaning that no quantum correction is necessary and the electron-positron pair production can be ignored. The calculation of the classical radiation parameter is also necessary when we consider ultrahigh intensities, and is given by:
\begin{equation}
 Rc = \frac{\alpha e E_0}{m w_0}{\chi}
\end{equation}
where $\alpha$ is the fine structure parameter and is approximately equal to $1/137.037$. In our model, $R_{c}$ is initially equal to 1.463$\cdot 10^{-4}$ at an maximum intensity of $I$ = \Wcm[5]{21}, which means that no radiation friction force should be considered. We calculated the two parameters $R_{c}$ and $\chi$ at each time step of the simulation to make sure that they don't exceed the value at which more physics should be taken into consideration. Also, by comparing the results obtained using the electron equation of motion incorporating radiation friction and those without radiation friction, we found that the effect of radiation friction force on electrons trajectories and energies is only significantly noticed at an intensity higher than $I$ $\geq$ \Wcm[5]{22}. Thus, the Newton-Lorentz equation describing the electrodynamics of an electron in an electromagnetic field can be simplified as follows:

\begin{equation}
\frac{d \textbf{p}}{d t}=- e(\textbf{E}+\textbf{v}\wedge \textbf{B})
\end{equation}

Under radial polarization, no longitudinal magnetic field component $B_z$ is formed, only a longitudinal electric field $E_z $ is present along the center of the focal spot and thus $B_z$= 0 resulting in equations governing electron energy and momentum of the following forms:
\begin{equation}
\frac{d p_x}{d t}=- e(E_x-v_zB_y)
\label{eq2}
\end{equation}
\begin{equation}
\frac{d p_y}{d t}=- e(E_y+v_zB_x)
\label{eq3}
\end{equation}
\begin{equation}
\frac{d p_z}{d t}=- e(E_z+v_xB_y-v_yB_x)
\label{eq4}
\end{equation}
\begin{equation}
m_0c^2\frac{d \gamma}{d t}=- e(v_xE_x+v_yE_y+v_zE_z)
\end{equation}
where , $ \gamma$, $m_0 $ are respectively the Gamma factor and the electron's rest mass. We use $ v_x=P_x/m_0\gamma$, $ v_y=P_y/m_0\gamma$ and $ v_z=P_z/m_0\gamma$ to rewrite eqs. \ref{eq2}-\ref{eq4}
\begin{equation}
\frac{d p_x}{d t}=- e(E_x-\frac{p_zB_y}{m_0\gamma})
\label{eq6}
\end{equation}
\begin{equation}
\frac{d p_y}{d t}=- e(E_y+\frac{p_zB_x}{m_0\gamma})
\label{eq7}
\end{equation}
\begin{equation}
\frac{d p_z}{d t}=- e(E_z+\frac{p_xB_y-p_yB_x}{m_0\gamma})
\label{eq8}
\end{equation}
The Lorentz Factor is defined as \(\gamma^2=1+(p_x^2+p_y^2+p_z^2)/(m_0^2 c^2 )\) and the kinetic energy of an accelerated electrons is given by: \((\gamma-1) m_0 c^2=0.51(\gamma-1)\) MeV.
Equations (6)-(8) are coupled ordinary differential equations which have been numerically solved using a three-dimensional simulation code based on 5th order Runga-Kutta method to obtain electrons momentum and trajectories. The code fully provides the trajectory, the momentum, and the fields that each electron experiences at every time step of the simulation.

\subsection{\textbf{Results and discussion}}
A radially polarized electromagnetic field is assumed as an incident field. The calculations have been performed with monochromatic plane wave having a wavelength of 800 nm, focused under an \textit{f}/\# = 5 unless stated otherwise. We consider the pulse duration to be $\tau_1$= 53 fs. The propagation of the focused laser pulse can be visualized by calculating the electric field components from negative to positive time delays. We assume that all simulations start at $t_s$= -53 fs and stop whenever the kinetic energy of electrons becomes constant. This energy is going to be designated as the final kinetic energy of electrons in the paper. 

\subsubsection{\textbf{Electron’s trajectories and maximum kinetic energy:} }\hspace*{\fill} \\

In this section, a particular interest is drawn towards the existence of electrons with trajectories in the opposite direction to the propagation. This phenomenon is a direct effect of the laser’s radial polarization. The kinetic energy of electrons is discussed for different laser wavelengths.

\begin{figure*}[hbt!]
\centering
\includegraphics[width=12.5cm,height=9cm]{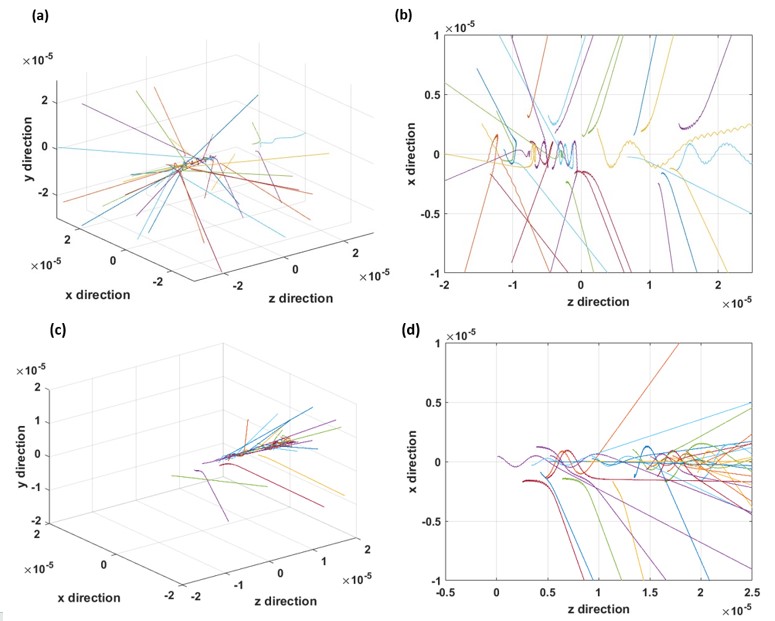}
\caption{\textsf{(a) 3D trajectories of electrons accelerated by a tightly focused and radially polarized laser pulse with an \textit{f}/\# = 5 and an intensity equal to $I$ = \Wcm[5]{19}. (b) Projection of the 3D trajectories in x-z plane, (c) 3D trajectories of electrons accelerated by a tightly focused and linearly polarized laser pulse with an \textit{f}/\# = 5 and an intensity equal to $I$ = \Wcm[5]{19}, (d) Projection of the 3D trajectories in x-z plane. (Unit here is in meters)}}
\label{fig:trajectories}
\end{figure*}
After they are born, the electrons gain a significant amount of momentum as a result of the high intensity of the laser. They are therefore pushed beyond the Rayleigh range and escape the laser with a speed close to the speed of light. On axis, the electrons experience only a longitudinal ponderomotive force equal to the longitudinal component of the electric field of the laser, since the radial component of the electric field and the azimuthal component of the magnetic field both vanish at all points on the propagation axis. Fig.\ref{fig:trajectories} (a) shows the 3D trajectories of accelerated electrons in both forward and backward directions under f/\# = 5 and an equivalent Gaussian laser focal intensity, $I$ = \Wcm[5]{19}. 
The backward electrons are shown to either start propagating conventionally (positive z direction) but end up changing direction and accelerate in the opposite direction to propagation as shown in Fig.\ref{fig:trajectories}  (b), or directly accelerate backwards right after they are born. This phenomenon has never been observed for linear (Fig.\ref{fig:trajectories}  (c) and (d)) and circular laser polarizations absent for Gaussian focusing effects. It is therefore clear that it is directly linked to the radial polarization that offers a unique characteristic once focused: a strong longitudinal electric field that becomes dominant when tightly focused \cite{Jeong2018}. 
Different characteristics of the incident radially polarized laser are shown to drastically increase the kinetic energy of the accelerated electrons \cite{Varin2013,Zaim2017,Salamin06-2} and GeV energies where reached \cite{Hu2006}. Here, we discuss the effect of the laser’s wavelength on the final energy of the accelerated electrons.
\begin{figure*}[hbt!]
\centering
\includegraphics[width=17cm,height=6cm]{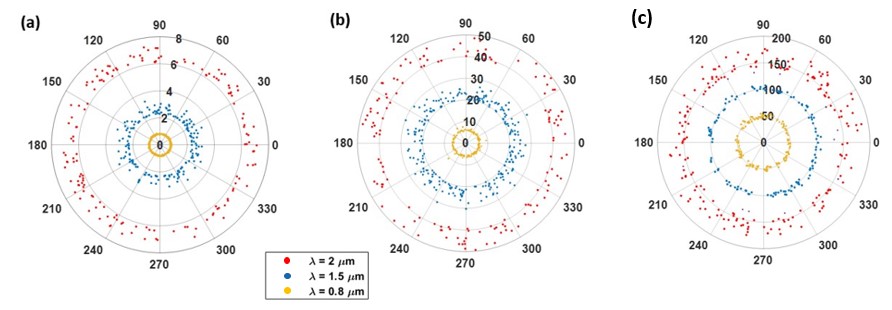}
\caption{\textsf{Polar plot of the final kinetic energy in MeV of accelerated electrons in x-y plane for different wavelengths and at different intensities.  (a)  $I$ = \Wcm[5]{19}, (b) $I$ = \Wcm[5]{20}, (c)  $I$ = \Wcm[5]{21}.}}
\label{fig:finalenergy}
\end{figure*}
 We use a polar plot to represent the final energy of the accelerated electrons in the x-y plane in Fig.\ref{fig:finalenergy}. The angle is defined as $\psi =\arctan(p_y/p_x)$ where $p_x$ and $p_y$ refer to the electron momentum in x and y directions respectively and $\psi$ is the azimuthal angle shown in Fig.\ref{fig:finalenergy} (b). The radius represents the final energy of the accelerated electrons. We use the function ‘atan2’ that returns the four-quadrant inverse tangent. Fig.\ref{fig:finalenergy} shows an important increase of the final energy of accelerated electrons for longer wavelengths. As one scans laser wavelength from $\lambda$ = 0.8~\um{} to $\lambda$ = 2~\um{}, the final energy of the accelerated electrons becomes two to three times larger depending on the incident laser intensity. The final energy can reach about 0.25 GeV at $I$ = \Wcm[5]{21} Gaussian for a 2 $\mu$m wavelength. High intensity lasers with longer wavelengths (up to 2 $\mu$m) offer an excellent way to accelerate electrons to GeV range. 
 
 \begin{figure*}[hbt!]
\centering
\includegraphics[width=15cm,height=9cm]{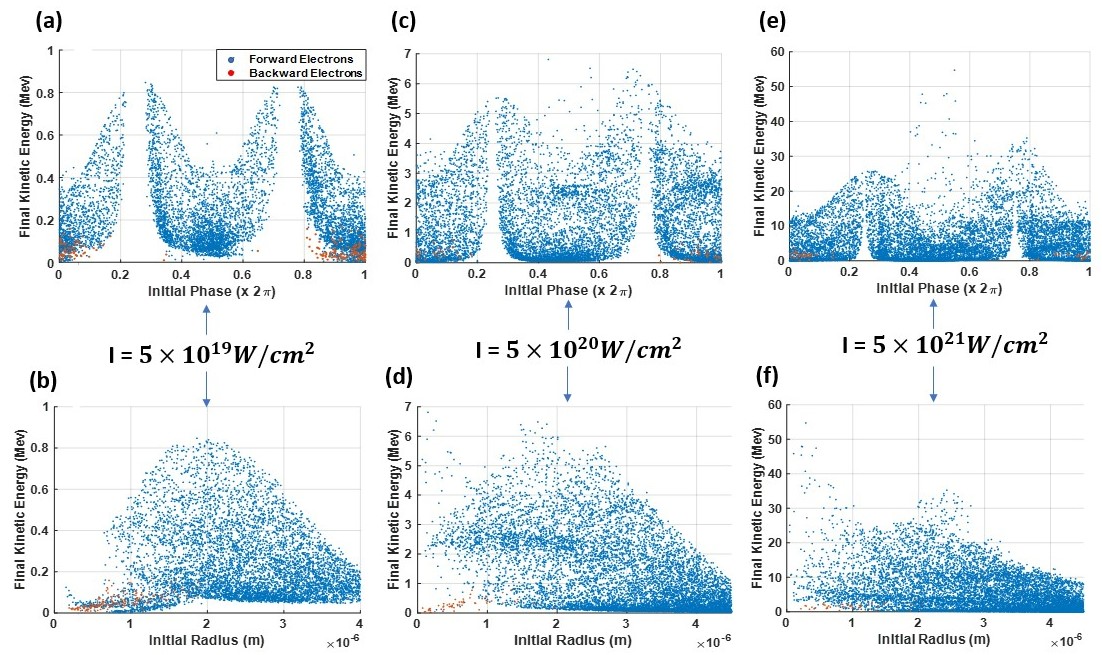}
\caption{\textsf{The initial phase and initial radius versus the final energy of accelerated electrons at an \textit{f}/\# = 5 and for different intensities, for $\lambda = 0.8 \mu$m (yellow polar plot in the previous figure).}}
\label{fig:fig6}
\end{figure*}
In order to elucidate the conditions necessary for maximum acceleration, we recorded the initial phase of the laser coincident with the birth of the electrons and the initial radii of said electrons that would be accelerated to high energies. First, Fig.\ref{fig:fig6} (a) shows that electrons born at phases close but not equal to $\pi/2$ and $3\pi/2$ accelerate to the highest kinetic energies. This peak of kinetic energy is given at a radius 1.4~\um{}~$\leq r \leq$~2.4~\um{} as shown in Fig.~\ref{fig:fig6} (b) which is the region where the radial component of the electric field is at its max. In Fig.\ref{fig:fig6} (c), in addition to initial phases close to π/2 and 3π/2 , the maximum kinetic energy is reached at initial phases around and equal to π, this phases match very small initial radii close to zero as shown in Fig.\ref{fig:fig6} (d). As the laser’s intensity increases, the peak kinetic energy moves to initial phases around and equal to π  as seen in Fig.\ref{fig:fig6} (e) , and very small initial radii as shown in Fig.\ref{fig:fig6} (f) which initially places the particle close to the \textit{z-}axis (propagation direction) where the longitudinal component of the laser’s electric field is at its max. The particles accelerated in the backwards direction have a lower final kinetic energy compared to the particles that accelerate conventionally: a low final kinetic value of 0.6 MeV is seen at $I$ = \Wcm[5]{19} and a max final kinetic value of less than 5 MeV is reached at an intensity $I$ = \Wcm[5]{21}. These unusual electrons are not born at the acceleration phase of the incident laser pulse as shown in Fig.\ref{fig:fig6} (a), (c), (e) and (g), and thus do not gain a considerable amount of energy. We can conclude that electrons gain a large amount of energy when they are born at the laser’s acceleration phase equal or close to π  and at a distance close to the propagation direction.

\subsubsection{\textbf{Radially polarized Electric field effect on the backward acceleration:} }\hspace*{\fill} \\

After showing the existence of particles propagating in the backwards direction, it is now imperative to understand the reasons behind it. The choice of the target radius is important to determine the electrons distance to the propagation direction and thus the axial electric field contribution on the backwards propagation. The effects of the radial electric field and the axial electric field components on the backwards propagation are then separately studied.

\begin{figure*}[hbt!]
\centering
\includegraphics[width=15.5cm,height=9.5cm]{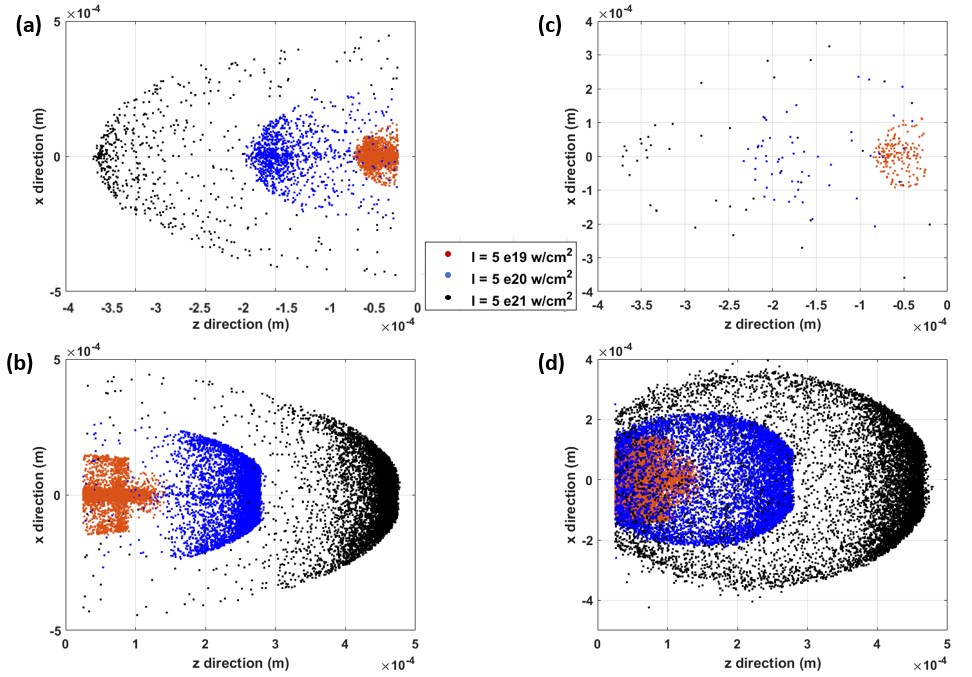}
\caption{\textsf{Final position of accelerated electrons in backward and forward directions at different intensities and different target radii. (a) Backward accelerated electrons from a cylindrical target with radius $R$ = $1$~\um{}, (b) Forward accelerated electrons from a cylindrical target with radius $R$ = $1$~\um{}, (c) Backward accelerated electrons from a cylindrical target with radius $R$ = $4$~\um{}, (d) Forward accelerated electrons from a cylindrical target with radius $R$ = $4$~\um{}}}
\label{fig:Fig7}
\end{figure*}
In Fig.\ref{fig:Fig7}, we considered two different targets with two radii $r =$~1~\um{} and $r=$~4~\um{} and with a length equal to the Raleigh range in both directions $-z$ and $z$. The f-number is equal to 5 and four different intensities are considered. Fig.\ref{fig:Fig7} (a) shows that more electrons are accelerated in the backward direction for a target with 1 micron radius while less are accelerated backwards for a larger radius of 4 microns Fig.\ref{fig:Fig7} (c). In Fig.\ref{fig:Fig7} (b) most of the accelerated particles travel far as they gain significant amount of energy and only few are left behind, while in Fig.\ref{fig:Fig7} (d) although most of electrons travel as far as in Fig.\ref{fig:Fig7} (b) a large population of them are left behind. A small target radius is favorable for accelerating electrons in the backward direction and helps the electrons in forward direction gain a considerable amount of energy and thus travel far in the propagation direction. This is mainly due to the strong longitudinal electric field in the region close to z-axis (propagation direction) which makes the electrons borne in this region gain a significant amount of energy and accelerate axially.
\begin{figure}
\centering
\colorbox{black!10}{\minipage[t][7cm][t]{8.5cm}
\hfill
\includegraphics[width=8.5cm,height=7cm]{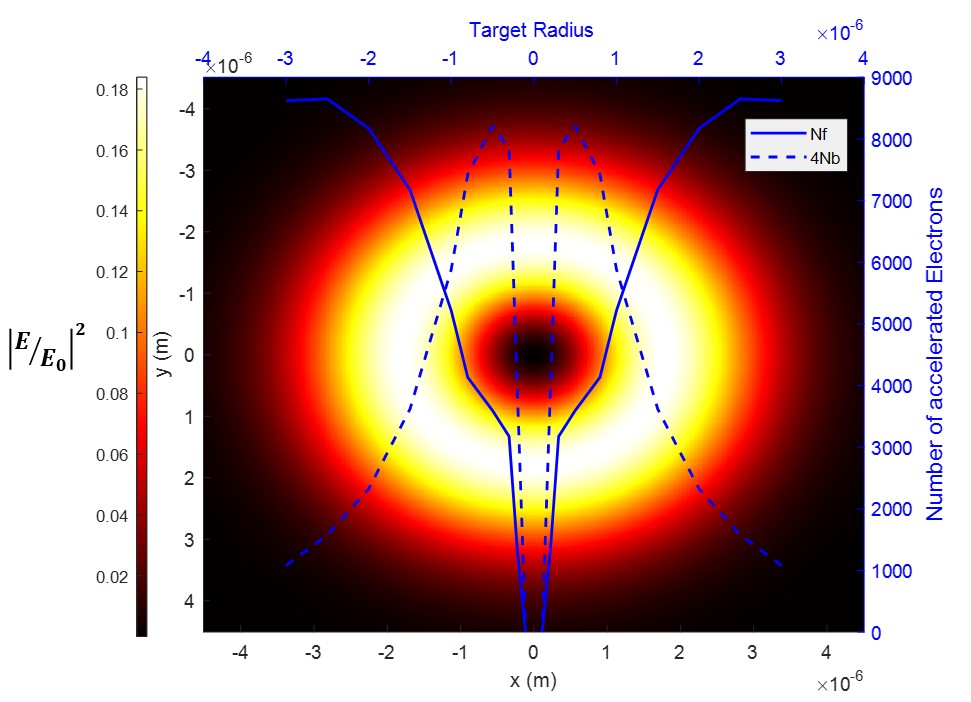}
\endminipage}
\caption{Distribution of the normalized squared electric field of the focal spot (\textit{f}/\# = 5 and $I$ = \Wcm[5]{19}) superimposed with the number of accelerated electrons versus the target radius (in blue). Nf is the number of electrons accelerated in the forward direction (Solid) and Nb is the number of electrons accelerated in the backward direction (Dashed).}
\label{fig:Fig8}
\end{figure}
We consider a hollow cylindrical target with a variable radius from 0 to 4 microns and a length equal to the Raleigh length. The neon ions are uniformly distributed on its surface. We calculate the number of particles accelerated on each direction for different values of the target radius. Fig.\ref{fig:Fig8} shows that the number of accelerated electrons in the forward direction drastically decreases as we get closer the center of the focal spot while the number of electrons accelerated in the backward direction increases as we get close to the null before it decreases at the zero-intensity region. This reveals that the electrons accelerated in the backward direction are mainly borne in the region of small radius (near to the \textit{z}-axis). None are seen on the propagation axis z because the radial electric field $E_r$  vanishes on axis and the longitudinal electric field $E_z$ is not strong enough to ionize Ne$^{7+}$. 
 \begin{figure*}[hbt!]
\centering
\includegraphics[width=15.5cm,height=6.5cm]{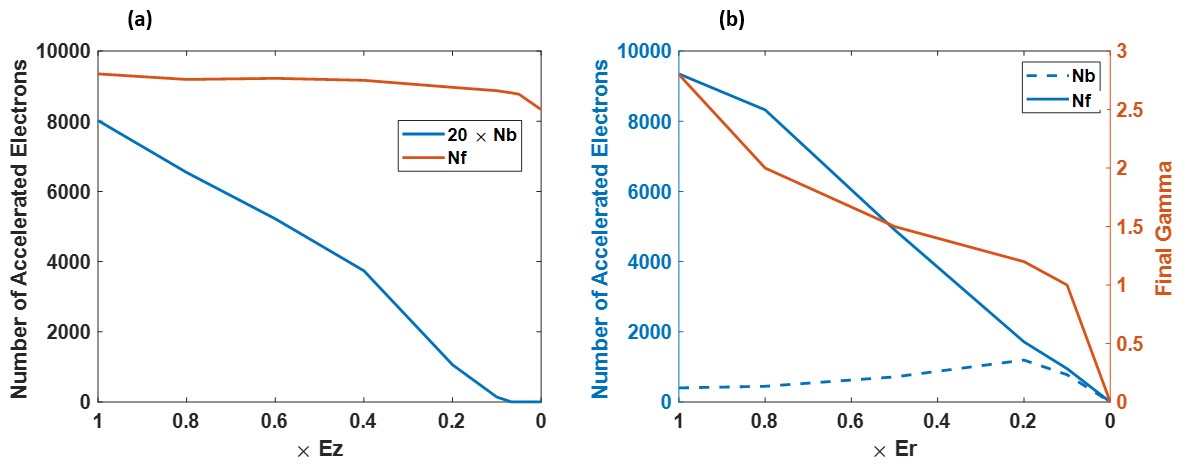}
\caption{\textsf{The number of accelerated electrons in backward (Nb) and forward (Nf) directions versus the artificially decreased longitudinal electric field$ E_z$ in a radially polarized laser pulse under an \textit{f}/\# = 5 and an intensity of $I$ = \Wcm[5]{19}. Here Nb is scaled by a factor of 20 in (a) and by a factor of 1 in (b)}}
\label{fig:Fig9}
\end{figure*}

\begin{figure*}[hbt!]
\centering
\includegraphics[width=15cm,height=12cm]{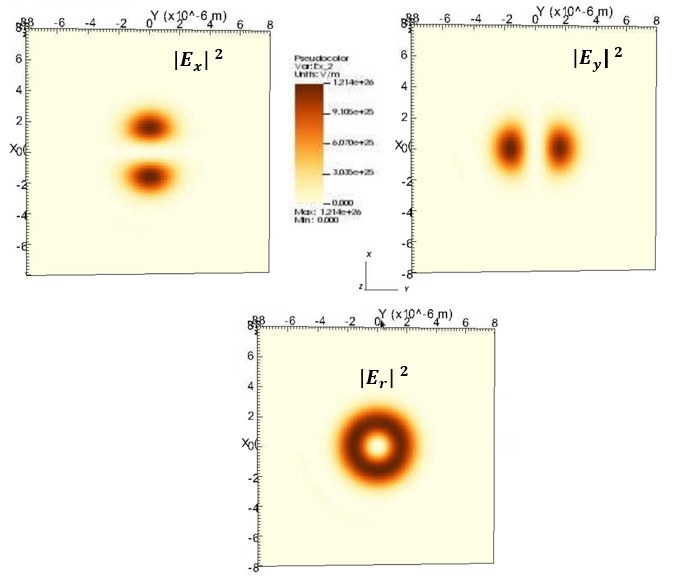}
\caption{\textsf{Projection of 3D PIC simulation: Orthogonal Hermite gaussian modes used to form the radial component of the electric field in radial polarization. (a)  representing the y component of the electric field squared , (b) TE 10 representing the z component of the electric field squared, (c) Superimposing (a) and (b) to get the doughnut shape characterizing radial component of the electric field squared }}
\label{fig:Fig10}
\end{figure*}

In order to understand how the incident laser’s electric field influences the generation and the acceleration of electrons in both forward and backward directions, we artificially decrease the strength of the longitudinal and the radial components of the electric field. Fig.\ref{fig:Fig9} (a) shows that the decreased longitudinal electric field $E_z$ leads to a significant decrease in the number of electrons accelerated in the backward direction. As $E_z$ approaches zero, no electrons are accelerated backwards. The particles accelerated in the forward direction slightly decreases but still exist for $E_z$=0. The final gamma of electrons does not decrease as electrons still accelerate in the forward direction independently of the decreasing $E_z$. Fig.\ref{fig:Fig9} (b) shows that the decreased radial electric field $E_r$ leads to an important decrease in the number of electrons accelerated in the forward direction, and thus a significant decrease of the final kinetic energy of electrons is observed. The number of electrons accelerated in the backward direction slightly increases as Er decreases before it drops to zero as Er becomes null. The longitudinal electric field component $E_z$ is responsible for accelerating electrons born at a close distance to z-axis in the backward direction, while $E_r$ is responsible about the energy gain and maintaining the acceleration in both directions.

\subsubsection{\textbf{PIC Simulation: Radial Polarization vs Linear Polarization} }\hspace*{\fill} \\
\begin{figure*}[hbt!]
\centering
\includegraphics[width=17cm,height=7.5cm]{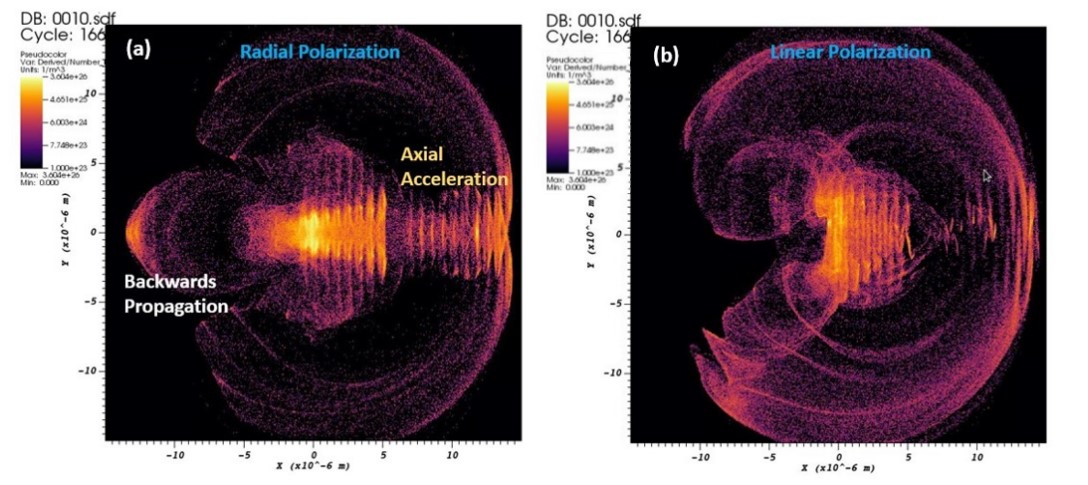}
\caption{\textsf{3D PIC simulation electron density snapshot at 100 fs: (a) Electrons density from ionization of $Ne^{8+}$ with radially polarized laser, (b) Electrons kinetic energy from ionization of $Ne^{8+}$ with linearly polarized laser.}}
\label{fig:Fig11}
\end{figure*}
\begin{figure*}[hbt!]
\centering
\includegraphics[width=17cm,height=7.5cm]{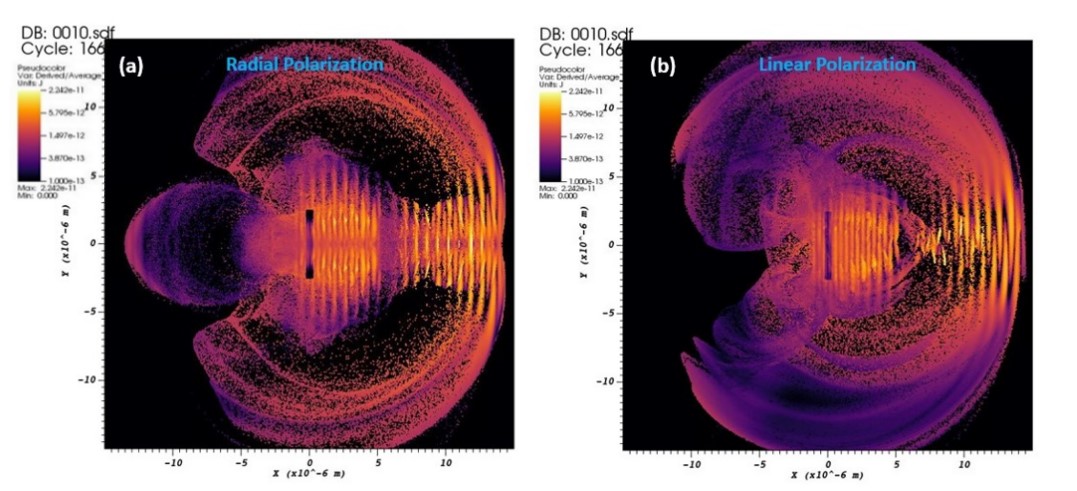}
\caption{\textsf{3D PIC simulation snapshot at 100 fs: (a) Electrons kinetic energy from ionization of $Ne^{8+}$ with radially polarized laser, (b) Electrons kinetic energy from ionization of $Ne^{8+}$ with linearly polarized laser.  Here in the color map the highest energy (yellow) and the lowest energy (purple/black) correspond to 140 MeV and 0.625 MeV, respectively.}}
\label{fig:Fig12}
\end{figure*}

Particle in Cell (PIC) codes are key tools for simulating and thus understanding laser-matter interaction in high energy density physics experiments \cite{Mouziouras2019,Smith2021}. The framework developed within EPOCH to model laser-plasma interaction at ultra-high intensities, provides a robust and computationally tractable mechanism for the generation and tracking electrons from atomic ionization\cite{Arber2015,Tuszewski2004}.
To complement our single particle simulation model presented previously, we run a 3D particle in cell simulation in EPOCH using a radially polarized laser with an intensity of $I$ = \Wcm[5]{21}. All other laser characteristics are similar to the Matlab single particle model. We then proceed to a comparison with the case of a linearly polarized laser of the same intensity and characteristics to show the existence of the backwards acceleration of electrons in the presence of radially polarized ultrahigh intensity laser.
Since Epoch only supports linearly polarized laser pulses in the standard laser model, we simulated our radially polarized laser by superimposing two orthogonal Hermite Gaussian modes\cite{Padgett2000,Pathak2016}, we then introduced a pi/2 phase delay to obtain the doughnut shape that characterizes the radial polarization as shown in Fig.\ref{fig:Fig10}.
The density of the ionized Neon target is as low as $1.10^{27}m^{-3}$ with a thickness of 0.5 $\mu$m and a radius of 5 $\mu$m. The propagation direction is set to \textit{x} and the laser hits the target at around 70 fs. The cell size is $50 \times 50 \times 50$ nm and the macro particles are considered to be 100 per cell.  Fig.\ref{fig:Fig11}-a shows electrons propagating axially in the backwards direction, and these electrons does not exist in the case of linear polarization (Fig.\ref{fig:Fig11}-b  which confirms the results given by the single particle code presented in the previous section. Also, a high density of electrons are shown in the axial forward direction due to the intense axial electric field Ex. In linear polarization case, the laser is y-polarized which means that the y component of the electric field is stronger than the it’s x(axial) component, that’s why Fig.\ref{fig:Fig11}-b shows less axially accelerated electrons compared to Fig.\ref{fig:Fig11}-a. Fig.\ref{fig:Fig12}-a shows that backwards propagating electrons carry less energy with a maximum of about 5.5 Mev compared to electrons propagating in the forwards direction with a maximum of 125 Mev. This again confirms the results obtained with the single particle simulation.

\subsubsection{\textbf{Backward Electrons: Initial conditions and propagation scenario:} }\hspace*{\fill} \\
\begin{figure*}[hbt!]
\centering
\includegraphics[width=17.5cm,height=5cm]{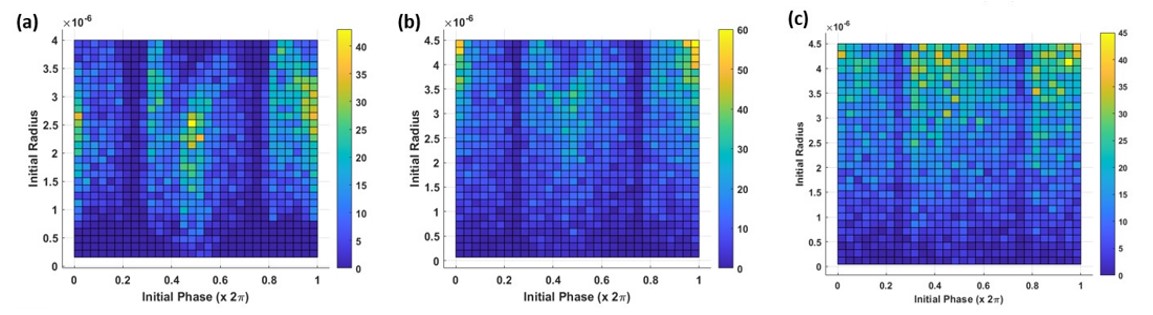}
\caption{\textsf{Projection of 3D histogram of the initial phases versus the initial radiuses of the accelerated electrons in the forward direction under an \textit{f}/\# = 5 and at different intensities. (a)  $I$ = \Wcm[5]{19}, (b) $I$ = \Wcm[5]{20}, (c)  $I$ = \Wcm[5]{21}.(Unit of y axis is meter) }}
\label{fig:Fig13}
\end{figure*}
\begin{figure*}[hbt!]
\centering
\includegraphics[width=17.5cm,height=5cm]{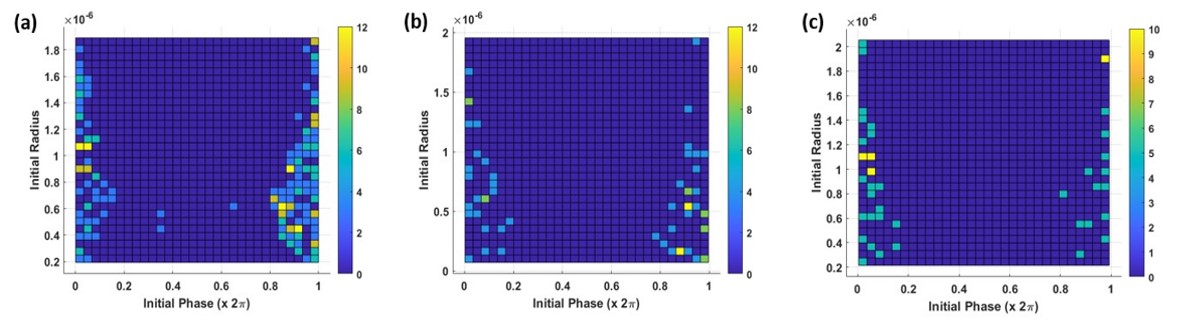}
\caption{\textsf{Projection of 3D histogram of the initial phases versus the initial radii of the accelerated electrons in the backward direction under an f-number of 5 and at different intensities.(a)  $I$ = \Wcm[5]{19}, (b) $I$ = \Wcm[5]{20}, (c)  $I$ = \Wcm[5]{21}. (Unit of y axis is meter)}}
\label{fig:Fig14}
\end{figure*}
To elucidate the particularity of electrons propagating in the backward direction, it is important to compare them to the conventional electrons that propagate in the forward direction. Since the initial conditions of an accelerated electron completely determine its trajectory and its direction of propagation, it is of prime importance to find out the initial phase and the initial radii at which the backward electrons are born. A backward propagation scenario is therefore elaborated and demonstrated to explain some of the backwards electrons behavior.
\begin{figure*}[hbt!]
\centering
\includegraphics[width=14cm,height=5.5cm]{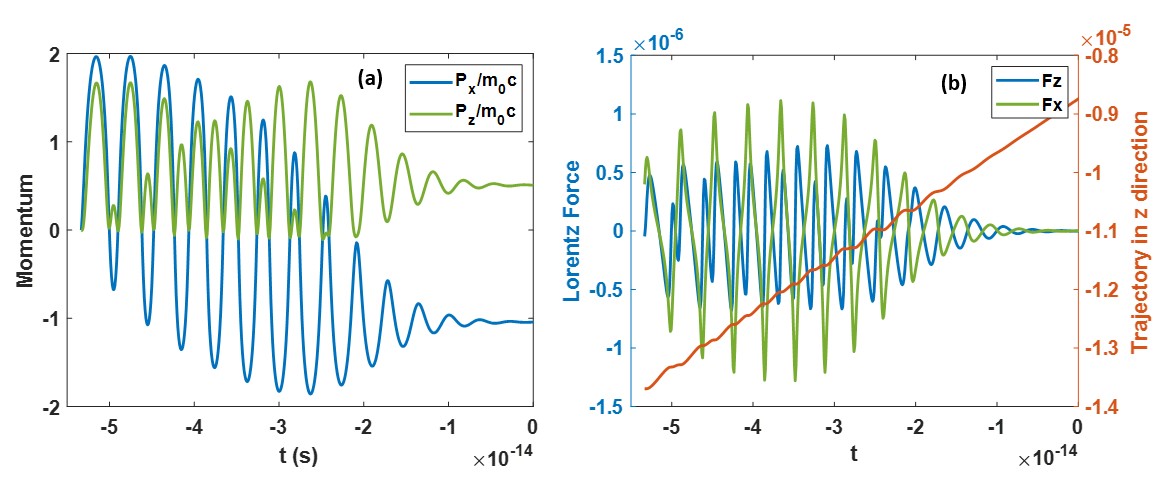}
\caption{\textsf{(a) Normalized momentums $P_x$ and $P_z$ of an electron accelerated in the forward direction under an f-number of 5 and an intensity of $I$ = \Wcm[5]{19}. (b) left: Lorentz force (SI units) components $F_x$ and $F_z$ of the same electron accelerated in the forward direction under an f-number of 5 and an intensity of $I$ = \Wcm[5]{19}, right: Electron trajectory in the propagation direction.}}
\label{fig:Fig15}
\end{figure*}
\begin{figure*}[hbt!]
\centering
\includegraphics[width=14cm,height=5.5cm]{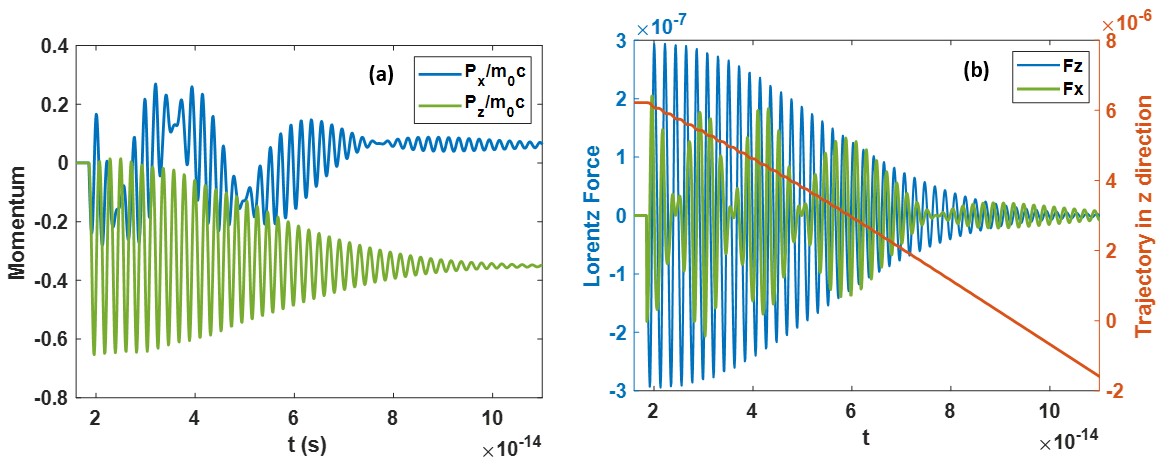}
\caption{\textsf{(a) Normalized momentums $P_x$ and $P_z$ of an electron accelerated in the backward direction under an f-number of 5 and an intensity of $I$ = \Wcm[5]{19}. (b) left: Lorentz force (SI units) components $F_x$ and $F_z$ of the same electron accelerated in the forward direction under an f-number of 5 and an intensity of $I$ = \Wcm[5]{19}, right: Electron trajectory in the propagation direction.}}
\label{fig:Fig16}
\end{figure*}
Fig.\ref{fig:Fig13} represents the histogram of accelerated electrons in terms of initial phase and initial radius and shows clearly that the forward acceleration is more favorable for large radii and phases around 0, $\pi$ and 2$\pi$. It is evident that no particles are accelerated at an initial phase of $\pi$⁄2 and/or 3$\pi$⁄2 for all intensities. Backwards acceleration is seen for smaller radii and for phases around zero (0 and 2$\pi$) as shown in Fig.\ref{fig:Fig14}. The backwards acceleration starts when the radius in smaller than 2 microns for $I$ = \Wcm[5]{19}, and this radius goes slightly larger as the intensity increases but still relatively smaller than the radius at which the forward propagation happens. At radial polarization, as the electron approaches nearer to the \textit{z-}axis (propagation direction), the longitudinal component of the laser’s electric field gets stronger making the particles born at a close distance to \textit{z-}axis (small radius) and at the right phase (around 0) accelerate in the backward direction.
To figure out the reason behind the backwards propagation and build a scenario that explains their behavior, a simplification of the equations is possible thanks to the symmetry of problem, and thus by considering y = 0, we can confine the electron dynamics to the propagation-polarization plane \textit{x-z} resulting in $p_y=0$ , $E_y=0$ and $B_x=0$. Therefore, the equations governing the electron motion become:
\begin{equation}
\frac{d p_x}{d t}=- e(E_x-\frac{p_zB_y}{m_0\gamma})
\label{eq11}
\end{equation}
\begin{equation}
\frac{d p_y}{d t}=0
\label{eq12}
\end{equation}
\begin{equation}
\frac{d p_z}{d t}=- e(E_z+\frac{p_xB_y}{m_0\gamma})
\label{eq13}
\end{equation}
Fig.\ref{fig:Fig15} (b) shows that initially the Lorentz force $P_z$ is positive giving the electron an initial push in the forward direction while in Fig.\ref{fig:Fig16} (b) the electron received a negative initial push which explains its propagation in the backward direction. If $E_z$ field is initially positive, the electron is accelerated backwards while when it is initially negative, the electron accelerates conventionally in the forward direction. This is the acceleration scenario for almost all electrons propagating in the backward direction. 
We can conclude that the initial direction of the $E_z$ field (pointing forward or backward) coupled with the initial phase and initial radius are extremely important in determining the acceleration direction of the electron. This unconventional phenomenon is only seen when a radially polarized laser pulse is tightly focused giving birth to a strong longitudinal field, otherwise, if $E_z$ is not strong enough Fig.\ref{fig:Fig9} (a), no electrons will be accelerated in the backward direction.

There are a few electrons with a different acceleration scheme as shown in Fig.\ref{fig:trajectories}. These electrons start accelerating in the forward direction before they suddenly change direction and continue their propagation in the backward direction. A mathematical model describing the interaction is necessary to fully understand the mechanisms behind the backward acceleration under radially polarized short pulse laser, and will be the focus of future work.

\subsection{\textbf{Conclusion}}
During the ionization of high charge state of Neon gas using a radially polarized and tightly focused petawatt class laser, electrons are directly accelerated in both forward and backward directions. The final kinetic energy of the particles accelerated in both directions can be increased for larger laser wavelengths. Near GeV energies were predicted at λ=2 microns and under an f-number of 5 and an intensity of $I$ = \Wcm[5]{22} Gaussian. Initial conditions (phase and radius) of the Ne8+ ions were determined for high kinetic energy. An interesting feature of the radial polarization is the formation of a strong longitudinal electric field along the propagation axis that becomes dominant when tightly focusing to few microns and/ or sub-micron focal spot. By artificially decreasing the strength of the axial electric field component $E_z$, it was shown that it is responsible about accelerating electrons born at a close distance to z-axis and at around 0 and 2π phases in the backward direction. The kinetic energy of this unusual electrons is low compared to the conventional electrons and this is because they are generally not born at the acceleration phase of the laser which is $\pi$. This unique feature is only seen when the incident laser pulse is radially polarized and tightly focused. The radial component of the electric field $E_r$ is shown to be responsible about the energy gain and therefore maintains the acceleration in both directions.

\section*{Acknowledgement}
This work was partially supported by the US Department of Energy Award \# DE-SC0020242 and US Air Force Office of Scientific Research award \# FA9550-22-1-0549. The authors also acknowledge support from the Ohio Supercomputer Center and the US Department of Energy supercomputing infrastructure, NERSC. 


\end{document}